\documentclass[preprint,aps,showpacs]{revtex4}
\usepackage{amsfonts}
\usepackage{amsmath}
\usepackage{amssymb}
\usepackage{graphicx}

\begin{document}

\title{ Orbital Kondo effect modulated by off-diagonal orbital interference}
\author{Jun Wen, Ju Peng, Baigeng Wang$^{*}$, D. Y. Xing}
\affiliation{National Laboratory of Solid State Microstructures and Department of
Physics, Nanjing University, Nanjing 210093, China}
\date{\today}

\begin{abstract}
We report the theoretical investigation of the orbital Kondo effect in an
Aharonov-Bohm interferometer by slave-boson mean field approach. It is found
that the present orbital Kondo effect can be tuned geometrically by the
external magnetic flux. When the magnetic flux $\varphi =(2n+1)\pi $, the
off-diagonal self-energy vanishes and the orbital Kondo problem can be
exactly mapped onto the usual spin Kondo model. For a general $\varphi $,
the presence of the off-diagonal orbital wave function interference will
modify the height and width of the orbital Kondo peak, but not change the
position of the orbital Kondo peak. We also give an analytic expression of
the flux-dependent Kondo temperature and find it decreases monotonously as
the magnetic flux $\varphi $ goes from $(2n+1)\pi $ to $2n\pi $, which means
the Kondo effect is suppressed by the off-diagonal orbital interference
process and becomes more easily destroyed by the thermal fluctuation. The
flux-dependence conductance is also presented.
\end{abstract}

\pacs{73.63.Kv, 72.10.Fk, 73.23.-b}
\maketitle

Very recently, quantum dots have attracted a considerable interest because
of their natural and potential applicability as basic blocks of solid-state
quantum computers. One of the important problems in quantum dots is the
Kondo problem, which exhibits the interplay between a localized spin of a
quantum dot and delocalized electrons in leads. As a result, the local spin
in quantum dot is screened by the coherent higher-order spin flip
co-tunneling process. Experimentally, a zero-bias peak of the differential
conductance with a width given by the Kondo temperature $T_{K}$ emerges. In
the last years, a great amount of works\cite%
{glazman,ng,meir,gordon,kouwenhoven,guo} was attributed to this problem.
Especially, Wilhelm \textit{et al}.\cite{wilhelm} proposed an original idea:
by defining two electrostatically coupled quantum dots, the orbital
structure of the wave functions acquires spin-like features, they can
exactly map the orbital-Kondo problem onto the spin-Kondo problem. Using the
four-terminal device, this idea was experimentally realized\cite{wilhelm2}
by the same group. Later, Lopez \textit{et al.}\cite{lopez} have discussed
the entanglement between charge and spin degrees of freedom when both
interdot and intradot Coulomb interactions exist. In this work, we will
consider the electron movement in a two-terminal Aharonov-Bohm interferometer%
\cite{yacoby,ji,iye,gefen,decker,buttiker,konig,claro,kang}, which has been
widely used to investigat the electron coherence. The schematic diagram of
the device is depicted in Fig.1. We suppose there are two quantum dots in
the different orbital paths of Aharonov-Bohm interferometer, and there
exists the Hubbard-type orbital interaction between two quantum dots. Note
that the present model is also related to the recent so-called pseudospin
Kondo correlation experiment\cite{holleitner}, in which the off-diagonal
orbital wave function interference information was neglected.

The model Hamiltonian we consider reads%
\begin{eqnarray}
H &=&\sum_{k\alpha =L,R}\varepsilon _{k\alpha }C_{_{k\alpha }}^{\dag
}C_{_{k\alpha }}+\sum_{i=u,d}\varepsilon _{i}d_{i}^{\dag }d_{i}+Un_{u}n_{d} 
\notag \\
&&+\sum_{k\alpha i}[T_{k\alpha i}C_{k\alpha }^{\dag }d_{i}+T_{k\alpha
i}^{\ast }d_{i}^{\dag }C_{k\alpha }].
\end{eqnarray}%
The first term stands for the Hamiltonians of the noninteraction leads $%
\alpha =L,R$ and $C_{_{k\alpha }}^{+}(C_{_{k\alpha }})$ are the
corresponding creation (annihilation) operators. The second term describes
the Hamiltonian of the up and down quantum dots, and $d_{i}^{+}(d_{i})$ are
the creation (annihilation) operators in the quantum dot $i$ with the
discrete energy level $\varepsilon _{i}.$ The third term is the interaction
between the up and down quantum dots$.$ The last one denotes the hopping
Hamiltonian between the lead $\alpha $ and dot with the hopping matrix
elements $T_{kLu,d}=t_{Lu,d}\exp (\mp \frac{i\varphi }{4}%
),T_{kRu,d}=t_{Ru,d}\exp (\pm \frac{i\varphi }{4})$. Here $\varphi $ is the
magnetic flux. Note that we have neglected the spin degree of freedom of
electrons, which contributes the simple factor $2$ in the electronic
current. We have also assumed that the hopping matrix elements $T_{kLu,d}$
are independent of the momentum index $k$ in the following calculation.

In order to deal with the inter-dot interaction, we use Coleman's
slave-boson mean field approach. Combining with Keldysh Green's function,
many authors\cite{langreth,schwab,lei,han,sun} have used slave-boson mean
field to study all kinds of transport problem. In the $U\rightarrow \infty $
limit, the localized electron operator $d_{i}$ can be replaced by $%
b^{+}f_{i} $ , where $b$ and $f_{i}$ being the standard boson and fermion
annihilation operators describing the empty $(n_{u}=0,n_{d}=0)$ and singly
occupied $(n_{u}=1,n_{d}=0)$ or $(n_{u}=0,n_{d}=1)$ states of up and down
quantum dots. Since the dots are either empty or singly occupied in the $%
U\rightarrow \infty $ limit, we have the following constriction condition%
\begin{equation}
b^{\dag }b+\sum_{i=u,d}f_{i}^{\dag }f_{i}=1.
\end{equation}%
Therefore, we can obtain the effective Hamiltonian in slave-boson
representation 
\begin{eqnarray}
H_{eff} &=&\sum_{k\alpha =L,R}\varepsilon _{k\alpha }C_{_{k\alpha }}^{\dag
}C_{_{k\alpha }}+\sum_{i=u,d}\varepsilon _{i}f_{i}^{\dag }f_{i}+  \notag \\
&&\sum_{k\alpha i}[T_{k\alpha i}C_{k\alpha }^{\dag }b^{\dag
}f_{i}+T_{k\alpha i}^{\ast }f_{i}^{\dag }bC_{k\alpha }]+  \notag \\
&&\lambda \lbrack \sum_{i}f_{i}^{\dag }f_{i}+b^{\dag }b-1].
\end{eqnarray}%
Note that we have incorporated the constriction condition into the effective
Hamiltonian by introducing a Lagrange multiplier $\lambda $. In the
mean-field approximation, $b$ and $b^{\dag }$ are replaced by a real c
number, $b=b^{\dag }=b_{0}.$ Because the slave-boson mean field is correct
for Kondo regime, we have to restrict our nonequilibrium calculation to the
low bias $V<<\mid \varepsilon _{i}\mid .$ With these preparations, the key
point for us is to determine the free parameters $\lambda $ and $b_{0}$
self-consistently by the condition%
\begin{equation}
\frac{\partial \langle H_{eff}\rangle }{\partial \lambda }=\frac{\partial
\langle H_{eff}\rangle }{\partial b}=0,
\end{equation}%
which gives the following two self-consistent equations%
\begin{equation}
-i\int \frac{dE}{2\pi }TrG^{<}(E)+b_{0}^{2}=1,
\end{equation}%
\begin{equation}
\lambda b_{0}^{2}=-\int \frac{dE}{2\pi }\sum_{\alpha }Tr\{G^{r}(E)\Gamma
_{\alpha }(E)f_{\alpha }(E)+\Gamma _{\alpha }(E)G^{<}(E)/2\}.
\end{equation}%
Once we have the parameters $b_{0}$ and $\lambda $, the present transport
problem becomes noninteraction one with the renormalized quantum levels and $%
\varepsilon _{i}\rightarrow $ $\varepsilon _{i}+\lambda $ and hopping matrix 
$T_{k\alpha i}\rightarrow b_{0}T_{k\alpha i}$. Further, by using the Keldysh
nonequilibrium Green's functions, the electronic current can be calculated
from the left lead, 
\begin{eqnarray}
I_{L} &=&q\langle \frac{d\hat{N}_{L}}{dt}\rangle  \notag \\
&=&-iq\int \frac{dE}{2\pi }Tr\{b_{0}^{2}\Gamma _{L}(E)\ast \lbrack
(G^{r}(E)-G^{a}(E))f_{L}(E)+G^{<}(E]\}.
\end{eqnarray}%
Where $f_{\alpha }(E)=\{\exp [E-\mu _{\alpha }]/k_{B}T+1\}^{-1}$ is the
Fermi distribution function and $[\Gamma (E)_{\alpha }]_{ij}=2\pi
\sum_{k}T_{k\alpha i}^{\ast }\ast T_{k\alpha j}\delta (E-\varepsilon
_{k\alpha })$ is the linewidth function. The Green's function $%
G_{ij}^{r,a,<}(E)$ is the Fourier transformation of $G_{ij}^{r,a,<}(t)$ with 
$G_{ij}^{r,a}(t)=\mp i\theta (\pm t)\langle \{f_{i}(t),f_{j}^{+}(0)\}\rangle 
$ and $G_{ij}^{<}(t)\equiv i\langle f_{j}^{+}(0)f_{i}(t)\rangle $.

In the following, we first calculate the retarded Green's function $G^{r}(E)$
by making use of the Dyson equation,%
\begin{equation}
G^{r}(E)=\frac{1}{E-H_{eff}-\Sigma ^{r}(E)},
\end{equation}%
where $\Sigma ^{r}(E)=\Sigma _{L}^{r}(E)+\Sigma _{R}^{r}(E)$ with%
\begin{equation*}
\Sigma _{L}^{r}(E)=-\frac{i\Gamma (E)b_{0}^{2}}{2}\left( 
\begin{tabular}{cc}
$1$ & $\exp (\frac{i\varphi }{2})$ \\ 
$\exp (\frac{-i\varphi }{2})$ & $1$%
\end{tabular}%
\right) ,
\end{equation*}

\bigskip 
\begin{equation*}
\Sigma _{R}^{r}(E)=-\frac{i\Gamma (E)b_{0}^{2}}{2}\left( 
\begin{tabular}{cc}
$1$ & $\exp (\frac{-i\varphi }{2})$ \\ 
$\exp (\frac{i\varphi }{2})$ & $1$%
\end{tabular}%
\right) .
\end{equation*}%
Where we have chosen the symmetric linewidth function $[\Gamma
_{L}(E)]_{ii}=[\Gamma _{R}(E)]_{ii}$ ; $[\Gamma _{L}(E)]_{ij}=[\Gamma
_{R}(E)]_{ij}^{\ast }$. Next, we solve the lesser Green's function $G^{<}(E)$
which is needed in the electric current formulism and self-consistent
equation. Usually, It is calculated from the Keldysh equation%
\begin{equation}
G^{<}(E)=G^{r}(E)\Sigma ^{<}(E)G^{a}(E)
\end{equation}%
with the lesser self-energy $\Sigma ^{<}(E)\equiv \Sigma _{L}^{<}(E)+\Sigma
_{R}^{<}(E)$, and 
\begin{equation*}
\Sigma _{L}^{<}(E)=i\Gamma (E)b_{0}^{2}\left( 
\begin{tabular}{cc}
$1$ & $\exp (\frac{i\varphi }{2})$ \\ 
$\exp (\frac{-i\varphi }{2})$ & $1$%
\end{tabular}%
\right) f_{L}(E),
\end{equation*}

\bigskip 
\begin{equation*}
\Sigma _{R}^{<}(E)=i\Gamma (E)b_{0}^{2}\left( 
\begin{tabular}{cc}
$1$ & $\exp (\frac{-i\varphi }{2})$ \\ 
$\exp (\frac{i\varphi }{2})$ & $1$%
\end{tabular}%
\right) f_{R}(E).
\end{equation*}%
Eqs.(5)-(9) constitute a closed solution to the present transport problem,
and we can calculate all kinds of physical quantities numerically. In the
following, we set the Fermi level of the leads be zero, $2\Gamma =1$ as the
energy unit and both quantum dots have the same energy level $\varepsilon
_{u}=\varepsilon _{d}=\varepsilon _{0\text{ }}<0.$ In addition, we assume
the square symmetric bands for leads, that is, $\Gamma (E)=\Gamma \theta
(W-\mid E\mid )$ with $W>>\mathrm{max}\{k_{B}T,\Gamma ,qV,\mid \varepsilon
_{i}\mid \}.$These parameters guarantee the system in the typical Kondo
regime.

In Fig.2 we plot the local density of state $(LDOS\equiv -\frac{\mathrm{Im}%
[G_{u}^{r}(E)]}{\pi }=-\frac{\mathrm{Im}[G_{d}^{r}(E)]}{\pi })$ versus the
energy with the different mgnetic flux $\varphi $ and bare energy level $%
\varepsilon _{0\text{ }}$. It is found that the $LDOS$ shows the well-known
Kondo peaks at $E=0$. To demonstrate the physics origin of the orbital Kondo
peak, we first discuss the special case with the magnetic flux $\varphi
=(2n+1)\pi $ [see also Fig.2(a)]. Since the off-diagonal retarded
self-energy [ $\Sigma _{L}^{r}(E)+\Sigma _{R}^{r}(E)$ ] becomes zero at
points $\varphi =(2n+1)\pi $, the orbital kondo problem is equivalent to the
usual spin Kondo one. The physics of this orbital Kondo resonance results
from the interesting co-tunneling process that is shown in Fig.3. When both
of the quantum dot energy level $\varepsilon _{i\text{ }}<0$ and there is
the strong Anderson interaction between two quantum dots, only one electron
is occupied in one dot (we assume the down dot ). Although the first-order
tunneling is blocked, the higher-order tunneling process still happens: the
first electron in the down quantum dot tunnels to the Fermi level of the
right lead via the down arm, the second electron at the Fermi level of the
left lead tunnels into the up dot via the up arm on a very short time scale $%
\sim \hbar /(\mu -\varepsilon _{0\text{ }}).$ Next, the second electron
repeats the process of the first electron via the up arm and another
electron tunnels into the down dot via the down arm. At low temperature, a
coherent superposition of all of this type co-tunneling process gives a
narrow Kondo resonance peak in the $LDOS$ of both up and down dots. The
position of the Kondo resonance is pinned at $E$ $=\tilde{\varepsilon}_{0%
\text{ }}\equiv (\varepsilon _{0\text{ }}+\lambda )\rightarrow 0$ and the
width is $\tilde{\Gamma}=\Gamma b_{0}^{2}.$ The Kondo temperature is
determined by $k_{B}T_{K}=\sqrt{\tilde{\varepsilon}_{0}^{2}+\tilde{\Gamma}%
^{2}}\sim W\exp (-\frac{\pi \mid \varepsilon _{0}\mid }{\Gamma }).$ In
addition, we also find two features in Fig.2(a-d): \ (1) For a given
magnetic flux $\varphi ,$ the width of Kondo resonance peak increases as the
bare energy level $\varepsilon _{0\text{ }}\rightarrow 0$. This is because
the empty state occupied probability increase when $\varepsilon _{0\text{ }}$
approaches to zero, and thus $b_{0}$ becomes larger. \ (2) For a given bare
energy level $\varepsilon _{0\text{ }}$, the position of the Kondo resonance
does not change with the magnetic flux. However, the width and height vary
with the magnetic flux $\varphi $. We can understand these as follows: For a
general $\varphi \neq (2n+1)\pi ,$ the off-diagonal orbital interference
results in the following diagonal retarded Green's function for up or down
quantum dot%
\begin{equation}
G^{r}(E)=\frac{E-\tilde{\varepsilon}_{0\text{ }}+i\tilde{\Gamma}}{(E-\tilde{%
\varepsilon}_{0\text{ }}+i\tilde{\Gamma})^{2}+\tilde{\Gamma}^{2}\cos ^{2}(%
\frac{\varphi }{2})},
\end{equation}%
which demonstrates that the position of the orbital Kondo resonance peak in $%
LDOS$ is still at $E=\tilde{\varepsilon}_{0\text{ }}\rightarrow 0,$ but the
height and width of the peak are modified by the factors $1/\sin ^{2}(\frac{%
\varphi }{2})$ and $\sin ^{2}(\frac{\varphi }{2})$, respectively. The
corresponding Kondo temperature can be obtained from%
\begin{equation}
k_{B}T_{K}(\varphi )=\sqrt{\tilde{\varepsilon}_{0}^{2}+\tilde{\Gamma}%
^{2}\sin ^{4}(\frac{\varphi }{2})}
\end{equation}%
We must point out that $T_{K}$ has a more complicated dependence of the
magnetic flux $\varphi $ due to $\tilde{\varepsilon}_{0}$ and $\tilde{\Gamma}
$ being still the implicit functions of $\varphi .$ In Fig.4, we demonstrate
the Kondo temperature $T_{K}(\varphi )$ decreases monotonously when the
magnetic flux $\varphi $ goes from $(2n+1)\pi $ to $2n\pi .$ This means that
the off-diagonal orbital wave function interference will suppress the
orbital Kondo effect. In particular, when the magnetic flux $\varphi $ is
close to $2n\pi $, the Kondo temperature of the system becomes very small
and the thermal fluctuation will more easily destroy the Kondo peak. Once
the temperature is higher than Kondo temperature $T_{K}(\varphi ),$ the
Kondo peak will suffer from a drift to the positions of the bare resonance.
In Fig.5, we plot the conductance as a function of the external magnetic
flux. It is found that although there are the same renormalized energy
levels $\tilde{\varepsilon}_{0\text{ }}$for both up and down quantum dots
which are very close to the Fermi energy of the leads, the destructive
interference results in a greatly suppressed conductance for a general
magnetic flux $\varphi $. Especially, the conductance vanishes when $\varphi
=(2n+1)\pi .$ This anti-resonance phenomenon has been discussed in reference%
\cite{konig}.

In summary, we have presented a flux-dependent orbital Kondo effect in an
Aharonov-Bohm interferometer. The off-diagonal orbital wave function
interference will modify the height and width of the orbital Kondo peak, but
not change the position of the orbital Kondo peak. The analytic result of
the flux-dependent Kondo temperature is also presented, and the numerical
calculations show the Kondo peak is robust to the wide range of the magnetic
flux $\varphi $. Since this present orbital Kondo effect can be tuned by the
geometric way, we hope this theoretical work will further stimulate the
experimental interests in the orbital Kondo physics.

\begin{acknowledgments}
This work was supported by the National Science Foundation of China under
Grant No. 90303011, 10474034 and 60390070. B. Wang was also supported by
National Basic Research Program of China through Grant No. 2004CB619305 and
NCET-04-0462.
\end{acknowledgments}

{$^{*}$ Corresponding author.}

Figure Captions

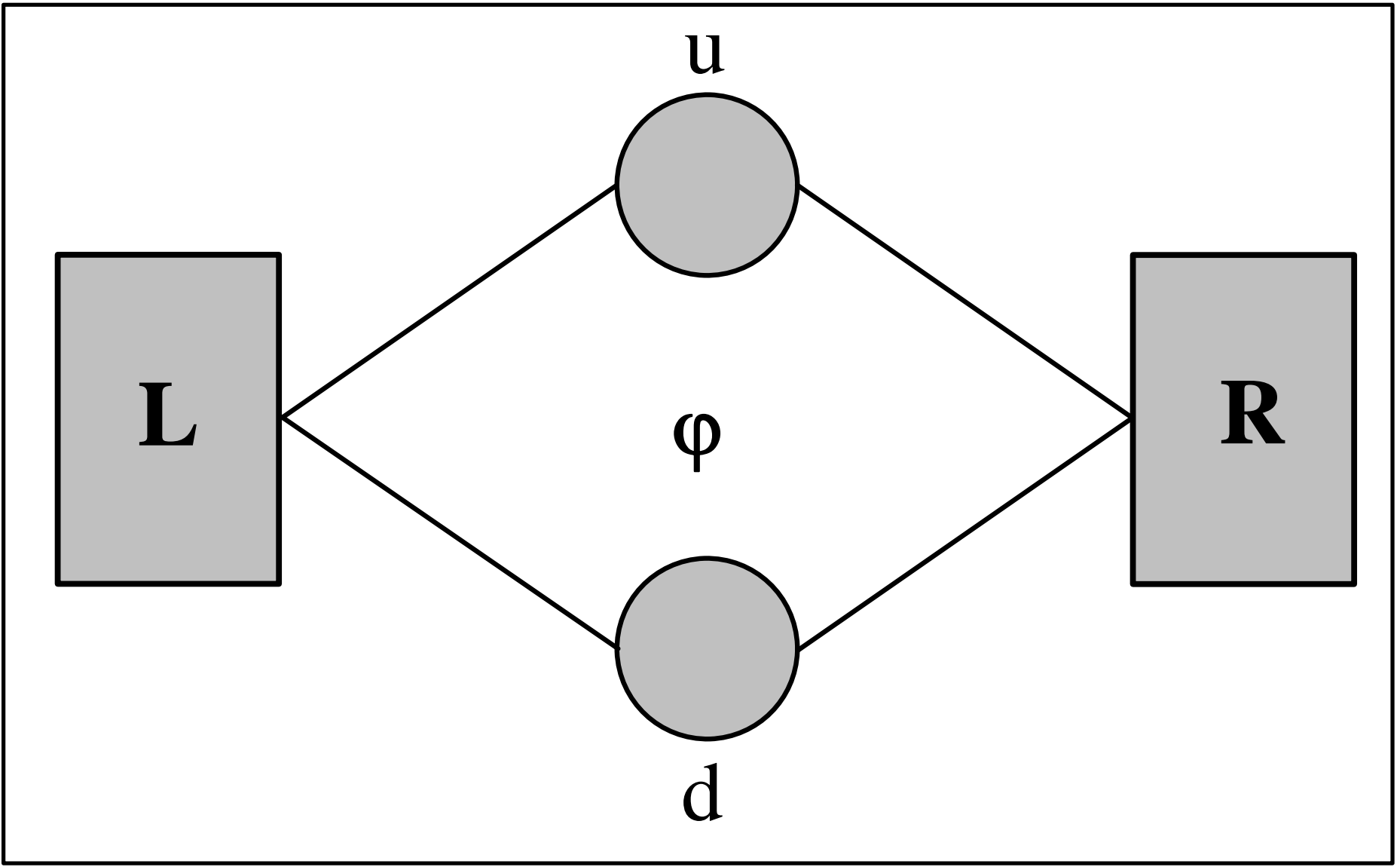
Fig. 1. Schematic diagram of our system.

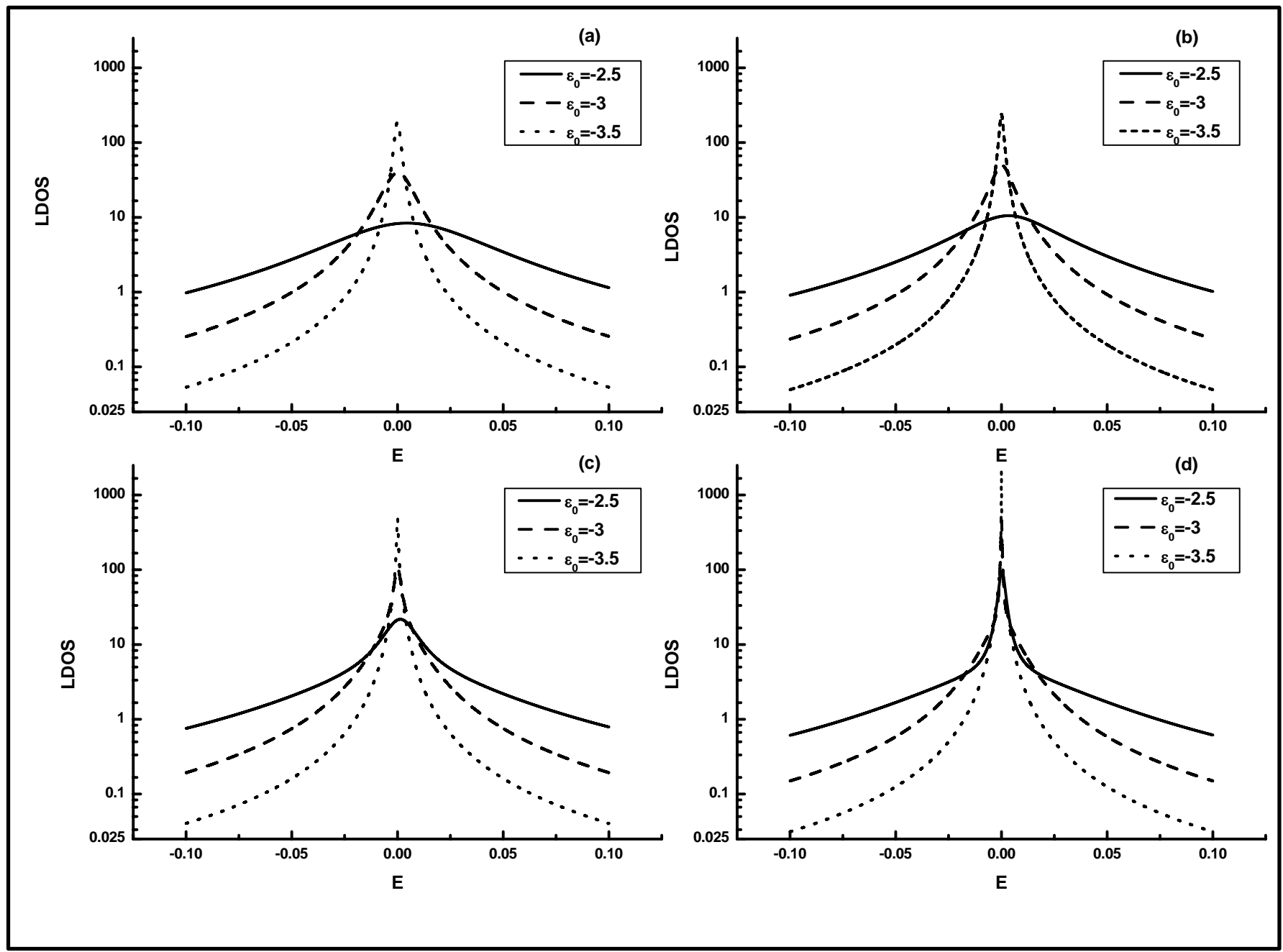
Fig. 2. The $LDOS$ of dependence of the energy for the different magnetic
flux $\varphi $: (a) $\varphi =\pi .$ (b) $\varphi =3\pi /4.$ (c) $\varphi
=\pi /2.$ (d) $\varphi =\pi /4.$ The solid, dashed and dotted lines in
(a)-(d) correspond to the different bare energy levels $\varepsilon _{0\text{
}}=-2.5,-3,-3.5,$ respectively. Other parameters are set: $2\Gamma
=1,W=100,\beta =10^{5}.$

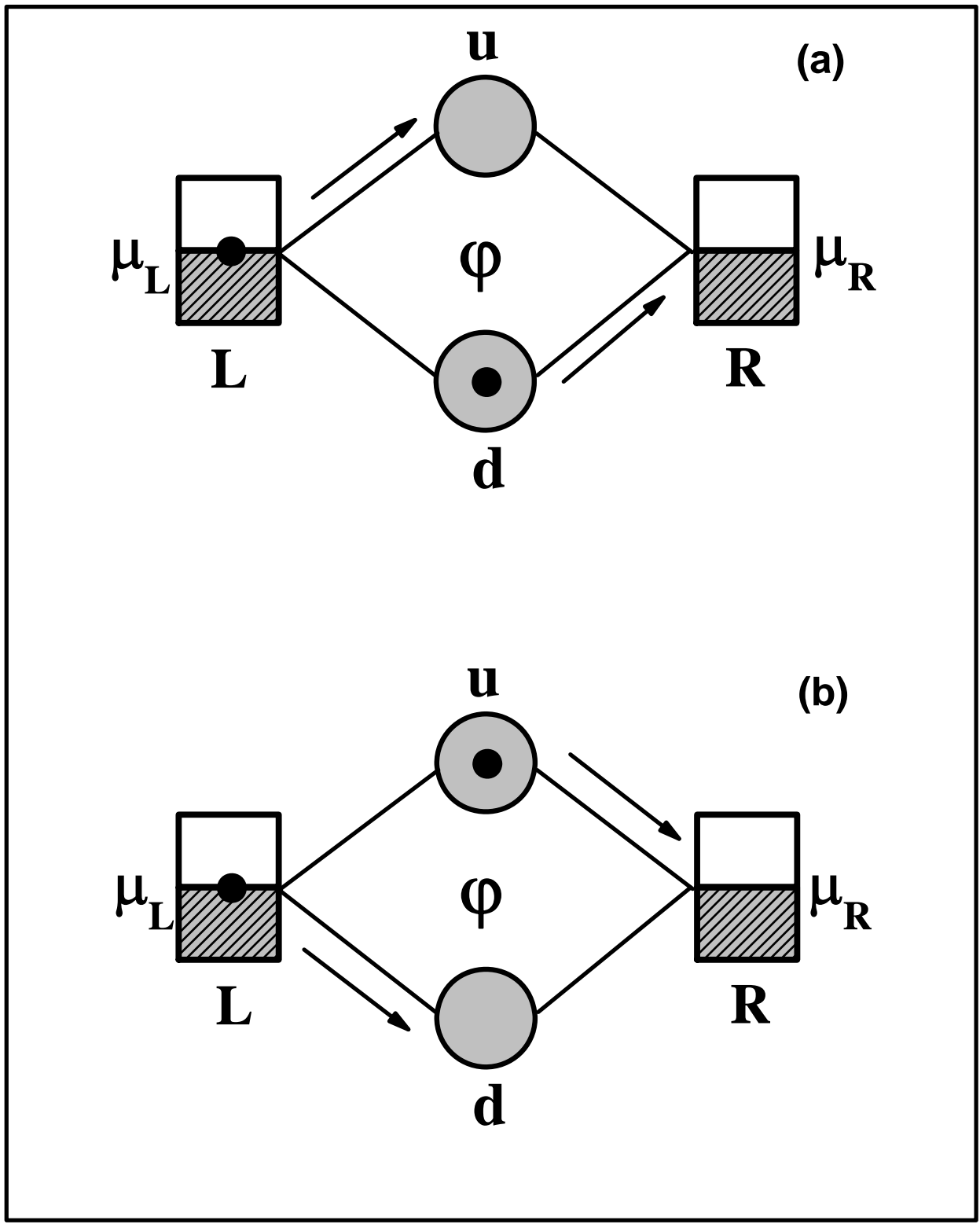
Fig. 3. The co-tunneling process giving rise to the orbital Kondo resonance:
(a) An electron in down dot tunnels into the right lead via the down AB arm,
followed by another electron in the left lead tunnels into the up dot via
the up AB arm. (b) An electron in up dot tunnels into the right lead via the
up AB\ arm, followed by another electron in the left lead tunnels into the
down dot via the down AB arm.

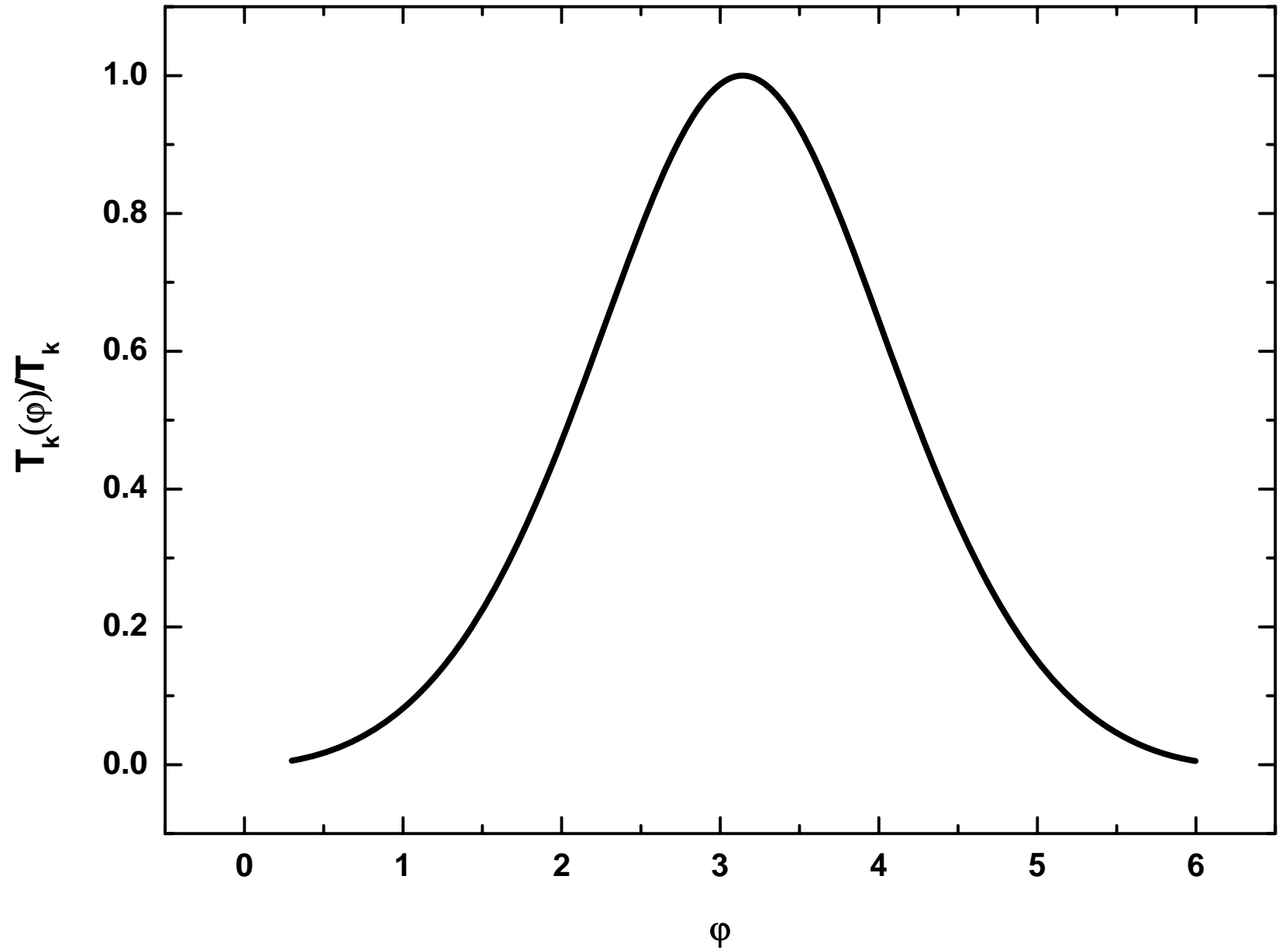
Fig. 4. Kondo temperature as the function of the magnetic flux $\varphi .$
We have set $\varepsilon _{0\text{ }}=-3.5$ and other parameters are the
same as those in Fig.2.

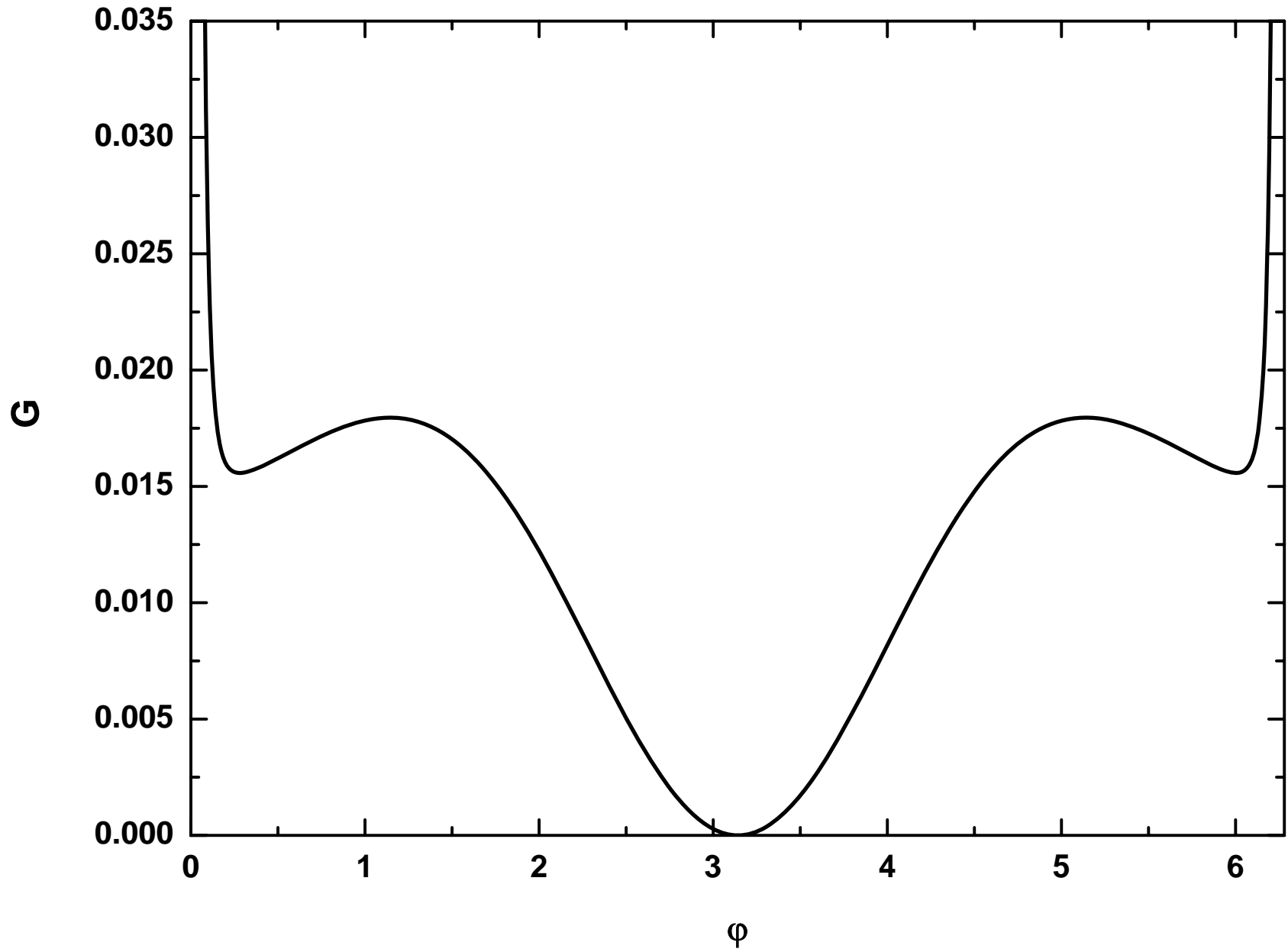
Fig. 5. The conductance versus the magnetic flux $\varphi .$ We have set $%
\varepsilon _{0\text{ }}=-2.5$ and other parameters are chosen as those in
Fig.2.

\end{document}